\documentclass{elsart}

\usepackage{graphicx}
\usepackage{amssymb}

\newcommand{\inte}{$INTEGRAL$}
\newcommand{\xmm}{$XMM-Newton$}

\def \hcm {\hbox {\ifmmode $ atom cm$^{-2}\else atom cm$^{-2}$\fi}}

\begin{document}

\begin{frontmatter}

% Title, authors and addresses

% use the thanksref command within \title, \author or \address for footnotes;
% use the corauthref command within \author for corresponding author footnotes;
% use the ead command for the email address,
% and the form \ead[url] for the home page:
% \title{Title\thanksref{label1}}
% \thanks[label1]{}
% \author{Name\corauthref{cor1}\thanksref{label2}}
% \ead{email address}
% \ead[url]{home page}
% \thanks[label2]{}
% \corauth[cor1]{}
% \address{Address\thanksref{label3}}
% \thanks[label3]{}

\title{Review on latest progress on \\
Supergiant Fast X--ray Transients \\
and future direction}

% use optional labels to link authors explicitly to addresses:
% \author[label1,label2]{}
% \address[label1]{}
% \address[label2]{}

\author{Lara Sidoli}
\ead{sidoli@iasf-milano.inaf.it}

\address{INAF/IASF-Milano, Via Bassini 15, 20133 Milano (Italy)}

\begin{abstract} 

In the recent years, the discovery of a new class of Galactic transients 
with fast and bright flaring X--ray activity, the Supergiant Fast
 X--ray Transients, has completely changed our view and 
comprehension of massive X--ray binaries.
These objects display X--ray outbursts which are difficult to be explained 
in the framework of standard theories for the accretion of 
matter onto compact objects, and could  represent a dominant population of X--ray binaries.
I will review their main observational properties 
(neutron star magnetic field, orbital and spin period, long term behavior, duty cycle, 
quiescence and outburst emission), which pose serious problems 
to the main mechanisms recently proposed to explain their X--ray behaviour.
I will discuss both present results and future perspectives with the next generation of X--ray telescopes.
\end{abstract}

\begin{keyword}
X--ray binaries, X--ray sources, accretion and accretion disks, supergiants
\end{keyword}

\end{frontmatter}

\section{Main observational properties of a new sub-class 
of High Mass X--ray Binaries: the Supergiant Fast X--ray Transients}

High Mass X--ray Binaries (HMXBs) contain a compact object (a neutron star or a black hole)
accreting matter from a massive  companion star. 
They can be divided into three different sub-classes, depending
on both the donor type (OB supergiant or Be star) and the X--ray activity (persistent
or transient): 
{\bf (1)}-supergiant HMXBs with persistent emission 
(divided into (1a)--wind-fed and (1b)--disk-fed accretors), 
{\bf (2)}-Be/X--ray transients (although there are also
a few Be/X--ray binaries with persistent low luminosity X--ray emission) and, more
recently, the {\bf  (3)}- supergiant HMXBs with fast transient emission, the so-called
Supergiant Fast X--ray Transients (SFXTs).
Liu et al. (2006) list
114  HMXBs located in our Galaxy, 66 of which are classified as accreting X--ray pulsars.
Most of them are in binaries with Be stars, although the number
of HMXBs with supergiant companions is continuously 
growing thanks to the \inte\ discoveries in
the energy range 17--100~keV: indeed, about 70\% of the HMXBs discovered with \inte\
host OB supergiants (Bird et al. 2010).

SFXTs are hard X--ray transients displaying a high dynamic range of $\sim$10$^{3}$--10$^{5}$, 
with sporadic, recurrent, bright and short 
(a few hour long) flares (Sguera et al. 2005, 2006;  Negueruela et al. 2006a), 
reaching 10$^{36}$--10$^{37}$~erg~s$^{-1}$.
This fast flaring activity is 
superimposed on outburst phases lasting a few days, shorter than those 
displayed by Be/X--ray transients
(Romano et al. 2007; Sidoli et al. 2009; Rampy et al. 2009).
Their optical association with blue supergiants 
has led to the identification of ten members 
(e.g. Halpern et al. 2004, Pellizza et al. 2006, Masetti et al. 2006, 
Negueruela et al. 2006b, Nespoli et al. 2008), together with 
several candidates with fast hard X--ray flaring activity but still  unknown 
optical/IR counterparts.

Their long term properties (on timescales of months) consist of a large flux variability
at an average intermediate X--ray luminosity of 10$^{33}$--10$^{34}$~erg~s$^{-1}$ (Sidoli et al. 2008), 
between the quiescence and the flare peaks.
The lowest luminosity level detected in a few SFXTs, 10$^{32}$~erg~s$^{-1}$, sometimes shows
a very soft spectrum (and, likely, no accretion; e.g. in IGR~J17544--2619, in't Zand 2005), sometimes 
a harder X--ray emission together with
mild flux variability (indicative of ongoing accretion at a very low level, e.g. IGR\,J08408--4503, 
Sidoli et al. 2010).
A common property of accreting pulsars in HMXBs is the 
X--ray spectral shape, typically characterized
by a flat hard power law below 10 keV (photon index $\sim$0--1), 
together with a high energy cut-off 
in the range 10--30~keV, sometimes
strongly absorbed at soft energies (Walter et al., 2006).
SFXTs display a similar spectral shape when they are in outburst, thus it is usually assumed that most
of these sources harbour neutron stars, although X--ray pulsations have been detected {\em only} 
in about  half of them (5 of about 10 members of the class)
with spin periods ranging from 4.7~s 
(AX~J1841.0--0536, Bamba et al. 2001) to 228~s (IGRJ~16465--4507, Lutovinov et al. 2005) and
1246~s (for the SFXT candidate IGRJ~16418--4532, Walter et al. 2006).
The possibility of the presence of a black hole in non-pulsating SFXTs cannot be completely ruled out.

%%%%%%%%%%%%%%%%%%%%%%%%%%%%%%%%%%%%%%%%%%%%%%%%%%
%%% 
\begin{figure}[ht!]
\begin{center}
\includegraphics*[angle=0,scale=0.58]{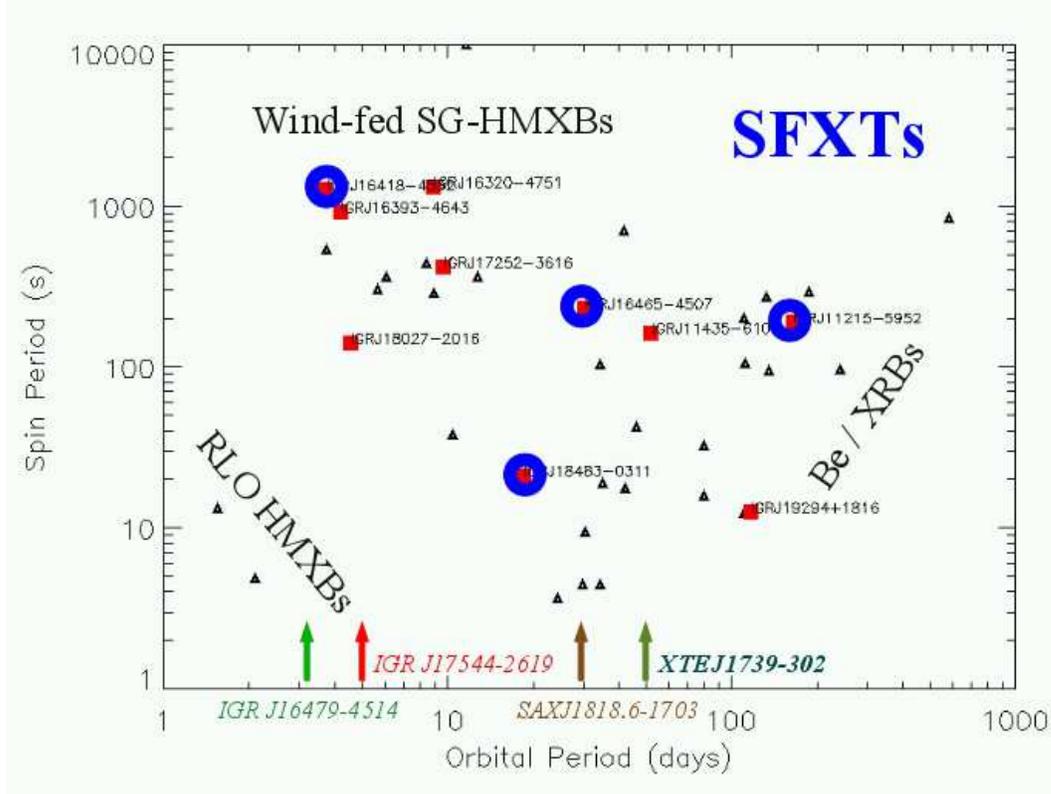} \\
\end{center}
\caption{\scriptsize Corbet diagram of Galactic accreting pulsars, together with
the new sources discovered with \inte\ (red squares), and a few SFXTs where both spin and orbital
periods are known (blue circles). On the orbital period axis, the position of other 
four SFXTs have been marked, for which the spin periodicities are still unknown.
}
\label{fig:corbet}
\end{figure}
%%%%%%%%%%%%%%%%%%%%%%%%%%%%%%%%%%%%%%%%%%%%%%%%%%

\begin{figure}[ht!]
\begin{center}
\includegraphics*[angle=0,scale=0.5]{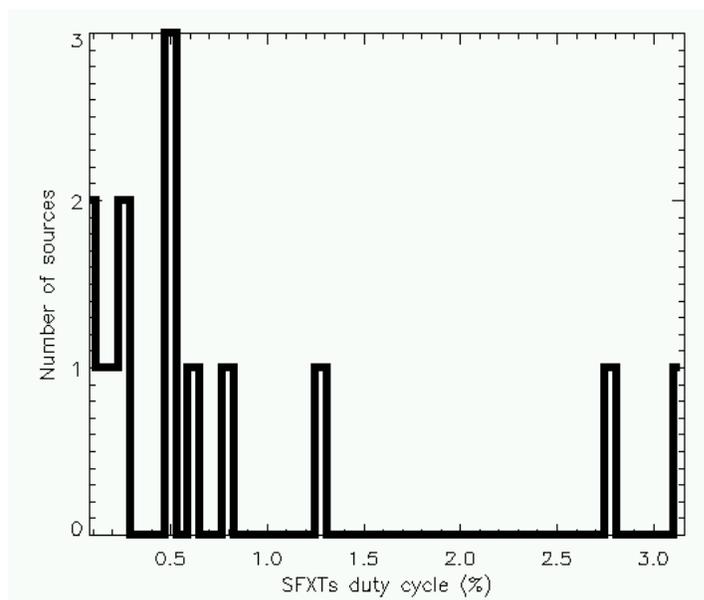} \\
\end{center}
\caption{\scriptsize Histogram of the SFXTs duty cycles 
(\% of time spent in bright flaring X--ray activity, 
as observed with \inte; data from Ducci, et al. 2010). 
}
\label{fig:dc}
\end{figure}

SFXTs orbital periods have been determined in 8 sources, spanning 
a large range as well, between 3.3~days (IGRJ~16479--4514; Jain et al. 2009) and 165~days 
(IGRJ~11215--5952; Sidoli et al. 2006, 2007; Romano et al. 2009).
The SFXTs for which both the orbital and spin periods are known can be overplotted in the
Corbet diagram of all known Galactic high mass X--ray pulsators 
(Fig.~\ref{fig:corbet}), where three different locii
were originally recognized (Corbet 1986): 
Be-star binaries (where spin periods are correlated with orbital periods),
supergiant systems with long spin periods ($\sim$100-1000~s; wind-fed HMXBs) and narrow orbits (P$_{orb}$$\sim$10~days) 
and supergiant systems with shorter spin periods ($\sim$1--10~s; Roche lobe overflow, disk-fed systems).
One of the puzzling facts about the new supergiant HMXBs discovered with \inte\
is that some SFXTs lie in the region typical for Be/X--ray transients (like IGR~J11215--5952)
or in an intermediate region of the Corbet diagram (like in the case of IGR~J18483--0311), 
in a sort of {\em bridge} between the two main locii of OB supergiants
and Be donors. 
It has been suggested that the SFXTs lying in the Be/XRBs region of the Corbet diagram
are indeed the descendant of these binary systems (Liu et al. 2010, Chaty 2010).
Interestingly, IGR~J18483--0311 shows also an X--ray flaring activity with a higher duty cycle (in excess of 3\%)
than typically observed from SFXTs (see Fig.~\ref{fig:dc}), possibly indicative of an intermediate
system between SFXTs and persistent HMXBs with supergiant companions, as suggested by Rahoui \& Chaty (2008).
The other source with a high hard X--ray duty cycle is IGR~J16479--4514 (2.7\%), with an unknown
spin period, while IGR~J11215--5952 has a small duty cycle of  0.3-0.6\%.
A sensitive Swift/XRT monitoring of IGR~J18483--0311 along an entire orbital period caught this
SFXT almost always active, except during a specific orbital phase, probably because of an X--ray 
eclipse or a gate mechanism (Romano et al. 2010).
The source light curve was highly variable, with several flares, with an average X--ray luminosity probably 
modulated
by the orbital period, likely because of an eccentric orbit (e$\sim$0.4, Romano et al. 2010).
Thus, it is also possible that a continuum of behavior exists between the persistent supergiant HMXBs and the
SFXTs with very rare X--ray activity. 
To date the SFXTs duty cycles 
observed at hard X--rays ($>20$~keV)
are typically small (Fig.\ref{fig:dc}) and highly variable from source to source.
Considering \inte\ observations of bright flaring activity, 
the percentage of time spent in flares with respect to the total observing time of the source field
can vary between 0.05\% and 3-4\% (for SFXTs in the central region of our Galaxy, after 7 years
of observations with \inte; Ducci et al., 2010).

%%%%%%%%%%%%%%%%%%%%%%%%%%%%%%%%%%%%%%%%%%%%%%%%%%
%%% 
%%%%%%%%%%%%%%%%%%%%%%%%%%%%%%%%%%%%%%%%%%%%%%%%%%

\section{Future perspective}

Despite the huge amount of observational data, several issues are still open:
one of the main problems is related with the link between SFXTs and 
HMXBs with supergiant companions and persistent X--ray emission.
%%%%%%
Indeed, these two kind of XRBs have both similar compact objects and donor stars, and
in a few cases, also very similar short orbital periods 
(as in the case of IGR~J16479--4514, Jain et al. 2009,
and IGR~J17544--2619, Clark et al. 2010).
%%%%%%
So, since in these particular cases, the possibility of different orbital parameters
are very likely ruled out as the main mechanism at the origin of the
two different classes, other mechanisms should be at work:
either a property of the neutron star (the magnetic field and/or the spin period,
as suggested by Grebenev \& Sunyaev 2007, Bozzo et al. 2008) or a property
of the OB supergiant: a different clumping factor in their strong winds
from the supergiant star in persistent HMXBs and in SFXTs, 
or the presence of a preferential plane
for the outflowing wind, inclined with respect to the orbital period in SFXTs,
which is crossed by the neutron star producing the 
SFXTs X--ray flaring activity (Sidoli et al. 2007).
%%%%%%
Also the ionization effect could play an important role especially in binary systems
with narrow orbits (Ducci et al. 2010).
%%%%%%
To date, none of the different mechanisms proposed for the transient
outbursts in SFXTs are able to explain all the observational facts (see Sidoli [2009] for a 
detailed review).

The possibility of a different accretion mechanism in persistent HMXBs with supergiant donors and in SFXTs
is also linked to the open issue of the evolutionary histories of these two subclasses of massive binaries.
Their census in our Galaxy, as well as their  proper duty cycles, are also open questions. 
%%%%%%
In conclusion, a more sensitive monitoring campaign of the hard X--ray sky is needed, 
at all timescales,  in order to try to answer to
all these questions,
to discover new periodicities (both spin and orbital periods) and know more about their nature, 
to accurately
model their broad band spectra on the shortest timescale possible, 
to possibly detect cyclotron lines and measure the neutron star magnetic field.
This latter is indeed unknown for {\em all} SFXTs, except in IGR~J18483$-$0311, where
a cyclotron emission line at 3.3~keV has been discovered with \xmm\ indicating
a low magnetic field of 3$\times10^{11}$~G (Sguera et al. 2010), thus ruling out
any magnetar nature for this SFXT.

In the following I will report on simulations of SFXTs
in the framework of two X--ray  missions perfectly suited
to study SFXTs and possibly solve most of their open issues:
the Energetic X-ray Imaging Survey Telescope (EXIST, Grindlay et al. 2010) 
and the New Hard X--ray Mission (NHXM, Tagliaferri et al. 2010).

\subsection{EXIST}

The high-energy telescope (HET, 5--600 keV) for EXIST was mainly suited to perform a survey of X--ray transients,
thanks to its large field of view (70 degrees $\times$ 90 degrees) and an 
unprecedented sensitivity (the full sky would have been observed 
in a two year continuous scanning survey), able to detect and quickly localize ($<$20$''$) X--ray sources for 
rapid follow-up.
Unfortunately, the Decadal Survey Report on the 13th of August 2010 did not recommend
EXIST. On the other hand, a future mission with the EXIST concept is certainly needed
to investigate the fast X--ray transient sky.

\begin{figure}[ht!]
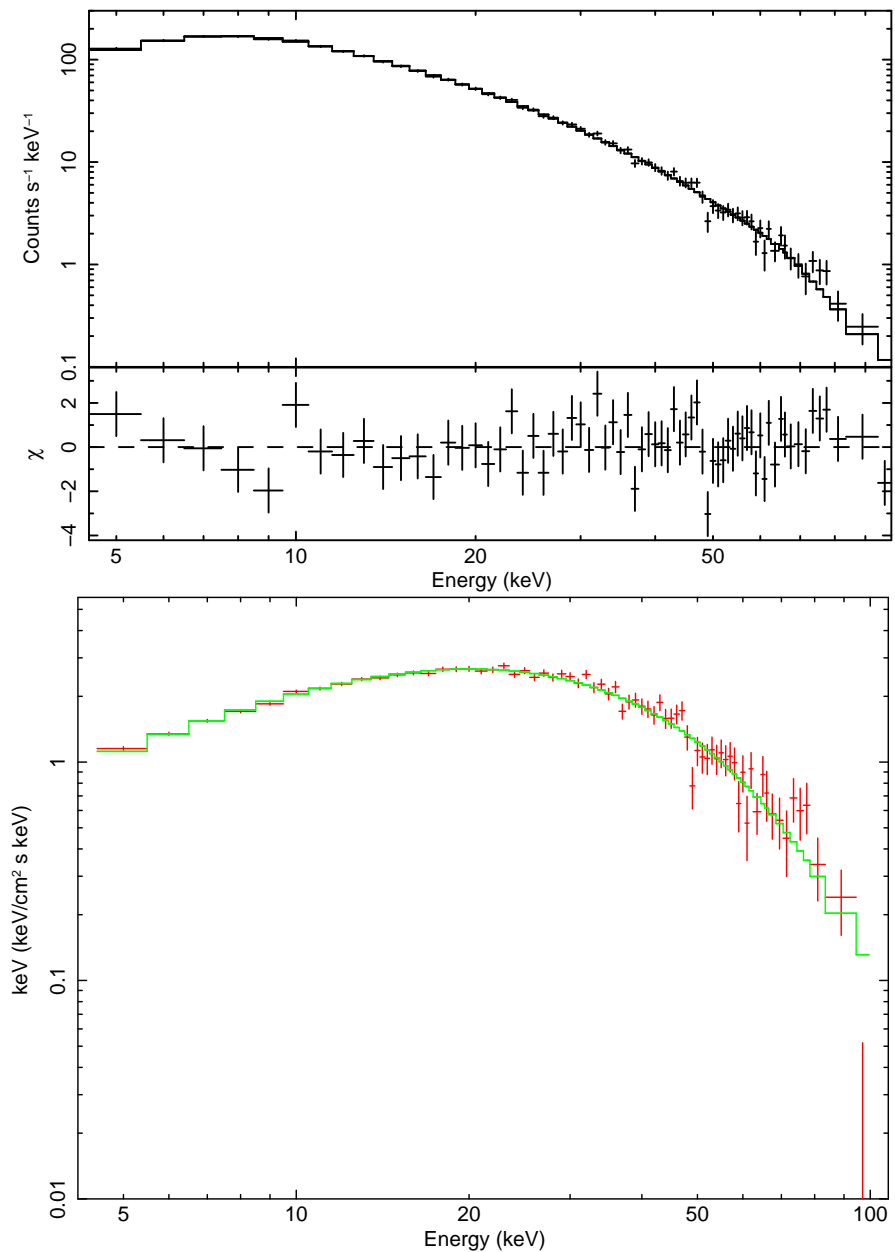

\begin{center}
\includegraphics*[angle=270,scale=0.5]{lsidoli_fig3a.ps}
\includegraphics*[angle=270,scale=0.505]{lsidoli_fig3b.ps}\\
\end{center}
\caption{\scriptsize Simulated spectrum  
with EXIST/HET (5--100 keV) of a SFXT in outburst (T$_{exp}$=1000~s). {\em Top panel}: Counts spectrum together
with the residuals, in units of standard deviations; {\em Bottom panel}: Energy spectrum (see text for the spectral parameters). 
}
\label{fig:exist_spec}
\end{figure}

In Fig.~\ref{fig:exist_spec} I report a spectrum simulated with HET,
with a spectral shape typical for a SFXT at the peak of its outburst,
with a short exposure time of 1~ks.
The spectral parameters are the following: 
a 5--100 keV flux of 8$\times$10$^{-9}$~erg~~cm$^{-2}$~s$^{-1}$, which translates into a net count
rate of 2360$\pm{9}$~s (HET, 5--100 keV), assuming a power law photon index of 0.66, a cut-off at 4~keV and
an e-folding energy at 15~keV.

%%%%%%%%%%%%%%%%%%%%%%%%%%%%%%%%%%%%%%%%%%%%%%%%%%
%%% 
%%%%%%%%%%%%%%%%%%%%%%%%%%%%%%%%%%%%%%%%%%%%%%%%%%

%%%%%%%%%%%%%%%%%%%%%%%%%%%%%%%%%%%%%%%%%%%%%%%%%%
%%% 
\begin{figure}[ht!]
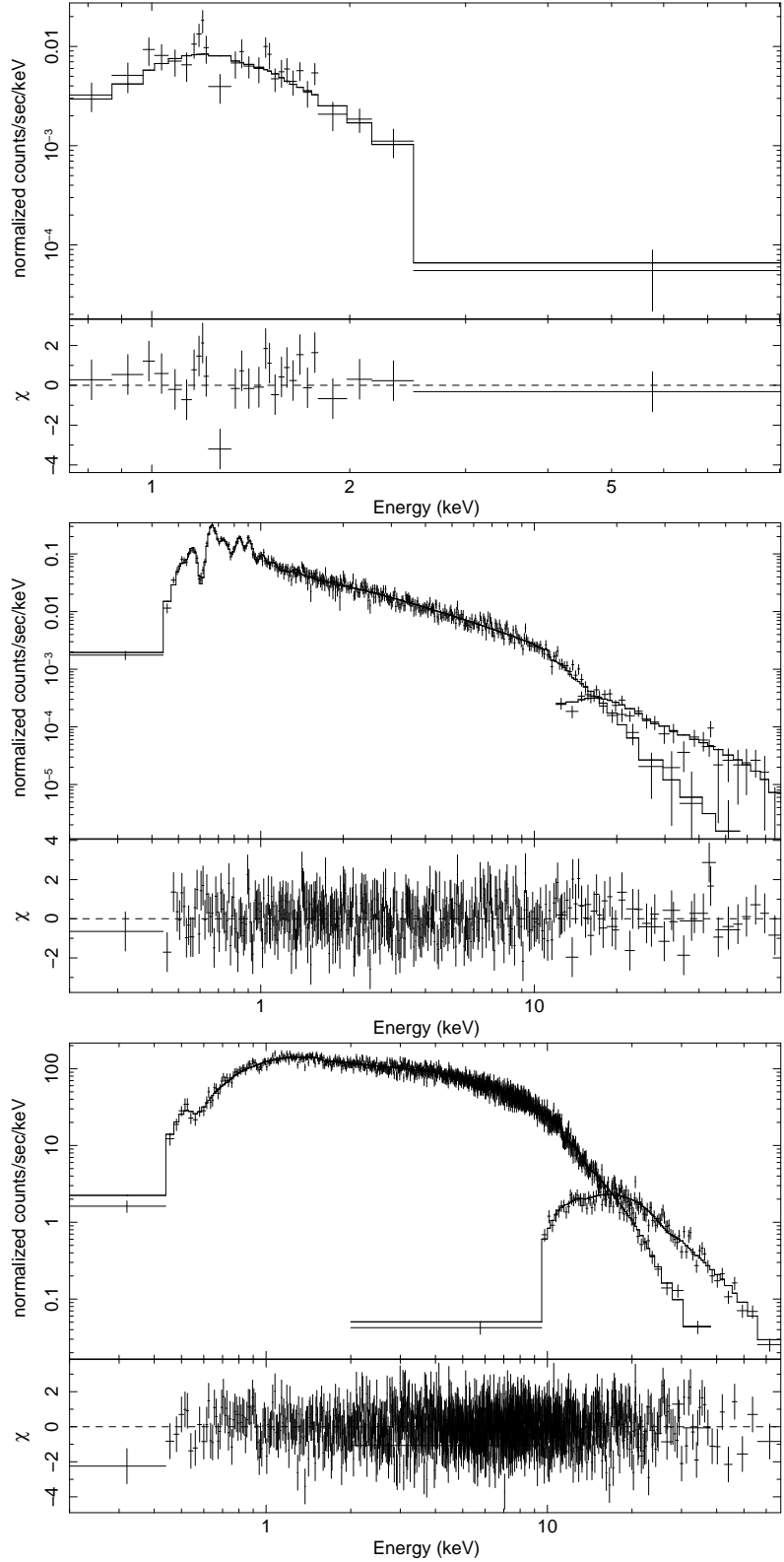

\begin{center}
\includegraphics*[angle=-90,scale=0.45]{lsidoli_fig4a.ps}
\includegraphics*[angle=-90,scale=0.45]{lsidoli_fig4b.ps}
\includegraphics*[angle=-90,scale=0.45]{lsidoli_fig4c.ps}\\
\end{center}
\caption{\scriptsize NHXM simulated broad band spectra of a SFXT in its three intensity states: 
in quiescence, with a flux
of 1.2$\times$10$^{-14}$~erg~cm$^{-2}$~s$^{-1}$ (1--10 keV; {\em Top panel}), 
in an intermediate
level of X--ray emission at 4$\times$10$^{-13}$~erg~cm$^{-2}$~s$^{-1}$ (1--10 keV; {\em Middle panel}),
and in bright flaring activity 
(8$\times$10$^{-9}$~erg~~cm$^{-2}$~s$^{-1}$, 1-100 keV; {\em Bottom panel}). 
See text for details on the spectral parameters. 
}
\label{fig:nhxm_spec}
\end{figure}
%%%%%%%%%%%%%%%%%%%%%%%%%%%%%%%%%%%%%%%%%%%%%%%%%%

%\vspace{3cm}
\clearpage
\subsection{NHXM}

The New Hard X--ray Mission (NHXM) is a medium size mission proposed for launch 
in 2016 and designed 
to gain important insights into several hot astrophysical issues, including 
the physics of accretion, the study of the highly absorbed and obscured sources,
and allowing to resolve the hard X--ray background.

NHXM will be able to obtain imaging and spectroscopic capability in the energy range 0.2--80 keV, 
together with sensitive photoelectric imaging polarimetry. 
It consists of four identical mirrors (the X--ray polarimeter is at the focus of one of these 
telescopes), 
with a 10 meters focal length, 
achieved thanks to a deployable structure. 
A low  background will be obtained thanks to a low equatorial orbit.
The field of view is of 12 arcmin, and a Half Power Diameter of 20 arcsec at 30 keV is expected
(Tagliaferri et al., 2010).

NHXM high sensitivity is particularly well suited for studying the fast spectral variability from SFXTs.
Indeed, their short timescales 
(from a few hundred second to a few hours) of the  
bright flaring activity have hampered to date a detailed investigation 
of their broad band spectral variability during short flares 
and an in-depth study of the variability of the absorbing material towards the line of sight, 
coming from the supergiant wind and directly accreting onto the compact object. 

A detailed modeling of the broad band spectrum is crucial in this respect, 
to well constrain the low energy absorption (and its variability) and to obtain 
physical spectral parameters (temperatures and optical depths of 
the Comptonizing plasma in the accretion column), 
instead of using simple phenomenological models (like power law with high energy cutoff).
This kind of investigation is essentially limited today by the source fast variability 
timescale and thus cannot be improved adopting huge exposure times with the present generation of instruments.

The spectral simulations I am reporting in Fig.~\ref{fig:nhxm_spec} 
demonstrate that NHXM will allow to probe the variability 
of the source properties on short timescales of a few hundred seconds during the outburst phase. 
In particular, simulations of SFXTs spectra adopting physical models 
(e.g. Comptonization emission, bulk motion Comptonization model, BMC) 
with a short exposure time of 100~s, allow already the determination of the 
spectral parameters with small uncertainties at a level of  few \%, 
to be compared with a 200~s exposure of the SFXT IGR~J08408$-$4503 with 
$Swift$/XRT and BAT, reported in Sidoli et al. (2009), where 
spectral parameters could be obtained only with an uncertainty 
at a level of 25\% (with a flux corrected for the absorption, in the 1--100 keV range, 
of 8$\times$10$^{-9}$~erg~~cm$^{-2}$~s$^{-1}$).

An open issue is also the measurement of the neutron star magnetic field 
in these sources by means of the detection of a cyclotron absorption line. 
NHXM, thanks to an accurate modeling of the broad-band continuum, 
will allow to observe a cyclotron absorption line at 12 keV already with an exposure time of 500~s, 
if the source is flaring at a flux of 8$\times$10$^{-9}$~erg~~cm$^{-2}$~s$^{-1}$.
Simulating a broad band continuum (hard power law with a high energy cutoff, 
similar to the IGR~J08408--4503, Sidoli et al. 2009) together with a cyclotron line 
at 12 keV (depth = 0.5; width =3 keV) and its harmonic (depth = 1; width = 5 keV), 
it is possible to obtain with NHXM (3 modules) the energy of the fundamental with an 
uncertainty of 0.8\% (90\% confidence level). 
Note that the energy range between 10 and 20 keV is 
hardly covered by the present generation of instruments.

The actual accretion mechanism can be investigated through the broad band spectroscopy, 
searching for changes in the absorbing column density. 
Some of these sources are highly absorbed, other are not, but still there is evidence 
that the absorbing column density is in excess of that towards the optical counterparts. 
This excess in the low energy absorption is very likely local to the sources and due to the supergiant wind. 
The possibility to probe the fast spectral variability on timescales of a few hundred
seconds is fundamental in unveiling the accretion mechanism, the structure of 
the supergiant wind (helping in disentangling anisotropic from spherically symmetric winds), 
and the interaction between the wind and the compact object. 

Another issue is the spectroscopy of the SFXTs intermediate intensity state 
(10$^{33}$--10$^{34}$~erg~s$^{-1}$) and of their quiescence level (10$^{32}$~erg~s$^{-1}$). 
In particular, a good quality spectrum of the true quiescence is lacking, 
and a study of the variability of both source states would be interesting, 
to better understand the nature of these objects. 

Observations with $Chandra$ and $XMM-Newton$ of a couple of SFXTs seem 
to indicate that their quiescent spectrum is very soft 
(power law photon index of 5.9, as in the $Chandra$ observation of IGR~J17544--2619, 
in't Zand 2005).
Preliminary simulations of this same model with NHXM demonstrate that, with an exposure time of 100~ks, 
it is possible to perform a good spectroscopy of the quiescent spectrum below 10 keV, 
finally distinguishing a featureless power law-like spectrum from a hot plasma model 
({\sc mekal} model in {\sc xspec}, see Fig.~\ref{fig:nhxm_spec}, top panel, 
F$_{X}$ (1--10 keV)=1.2$\times$10$^{-14}$~erg~cm$^{-2}$~s$^{-1}$, N$_{H}$=1.4$\times$10$^{22}$~cm$^{-2}$).
The spectroscopy of SFXTs at hard energies (above 20 keV) is possible to date only during bright outbursts. 
NHXM will allow to perform spectroscopy of SFXTs also during their intermediate intensity state 
(see Fig.~\ref{fig:nhxm_spec},  middle panel, 
for a simulation of a SFXT with the same spectral shape observed 
with $Suzaku$ from IGR~J08408--4503 by Sidoli et al. (2010), 
with an exposure time of 100~ks at a flux level of F$_{X}$ (1--10 keV)=4$\times$10$^{-13}$~erg~cm$^{-2}$~s$^{-1}$).

In conclusion, NHXM sensitivity will allow to really follow the evolution 
of both the spectrum and the absorbing column density in SFXTs, 
probing directly both the accretion process and the structure 
of the supergiant wind, during each single bright flare (a few hour long). 
These issues are expected to remain unsolved for years, since these questions need 
to be answered by a much higher sensitivity then the present instrumentation, 
to follow the rapid source variability.
   
\vspace{1cm}

%%%%%%%%%%%%%%%%%%%%%%%%%%%%%%%%%%%%%%%%%%%%%%%%%%%%%%%%%%%%%%%%%%%%%%%%
\normalsize
\textbf{Acknowledgements}\\
\scriptsize
I would like to thank the organizers of the event E16,  A.~Bazzano and N.~Gehrels, 
for their kind invitation at the COSPAR 2010 the 38th scientific assembly of
the committee on space research, held in Bremen on 18-25 July 2010.
I thank J. Grindlay for providing me with the EXIST/HET response matrix 
and L. Natalucci for his help with the EXIST simulations.
Thanks to F. Fiore, G.~Matt and G.~Tagliaferri for the NHXM response matrices.
This work is supported by contracts ASI/INAF I/008/07/0, I/033/10/0, I/088/06/0, I/069/09/0
 and partially funded by the grant from PRIN-INAF 2009, ``The
transient X-ray sky: new classes of X--ray binaries containing neutron stars''
(PI: L. Sidoli).

%%%%%%%%%%%%%%%%%%%%%%%%%%%%%%%%%%%%%%%%%%%%%%%%%%%%%%%%%%%%%%%%%%%%%%%%% 

\vspace{1cm}
\normalsize
\textbf{References}\\
\scriptsize
Bamba, A., Yokogawa, J., Ueno, M., et al., 2001, PASJ, 53, 1179-1183 \\
Bird, A.J., Bazzano, A., Bassani, L., et al., 2010, ApJS, 186, 1-9 \\
Bozzo, E., Falanga, M., Stella, L., 2008, ApJ, 683, 1031-1044 \\
Chaty, S., 2010, Proceedings of the Conference on 
``Binary Star Evolution: Mass Loss, Accretion, and Mergers, Mykonos'' (arXiv:1008.2718) \\
Clark, D.J., Hill, A.B., Bird, A.J. et al., 2009, MNRAS, 399, L113-L117. \\
Corbet, R. H. D., 1986, MNRAS, 220, 1047-1056 \\
Ducci, L., Sidoli, L., Paizis, A., 2010, MNRAS, 408, 1540-1550 \\
Grebenev,  S.A., Sunyaev, R.A., 2007, AstL, 33, 149-158 \\
Grindlay, J., Gehrels, N., Bloom, J., et al., 2010, SPIE, 7732, pp. 77321X-77321X-19 (arXiv:1008.3394) \\
Halpern, J.P., Gotthelf, E.V., Helfand, D.J., et al. 2004, The Astronomer's Telegram, 289 \\
in't Zand, J.J.M., 2005, A\&A, 441, L1-L4  \\
Jain, C., Paul, B., Dutta, A., 2009, MNRAS, 397, L11-L15 \\
Liu, Q.Z., van Paradijs, J., van den Heuvel, E.P.J., 2006, A\&A, 455, 1165-1168 \\
Liu, Q., Li, H., and Yan, J., 2010, Science in China G: Physics and Astronomy, 53, 130-134 \\
Lutovinov, A., Revnivtsev, M., Gilfanov, M., et al., 2005, A\&A, 444, 821-829  \\
Masetti, N., Morelli, L., Palazzi, E., et al. 2006,  A\&A, 459, 21-30 \\
Negueruela, I., Smith, D.M., Reig, P., et al. 2006a, in ESA Special Publication, ed. A.Wilson, Vol. 604, 165-170 \\
Negueruela, I., Smith, D.M., Harrison, T.E., et al., 2006b, ApJ, 638, 982-986  \\
Nespoli, E., Fabregat J., Mennickent, R.E., 2008, A\&A, 486, 911-917 \\
Pellizza, L.J., Chaty, S., Negueruela, I., 2006, A\&A, 455, 653-658 \\
Rahoui, F., Chaty, S., 2008, A\&A, 492, 163-166  \\
Rampy, R.~A., Smith, D.~M. \& Negueruela, I., 2009, ApJ, 707, 243-249  \\
Romano, P., Sidoli, L., Mangano, V. et al., 2007, A\&A, 469, L5-L8   \\
Romano, P.; Sidoli, L.; Cusumano, G.; et al., 2009, ApJ, 696, 2068-2074 \\
Romano, P., Sidoli, L., Ducci, L., et al., 2010, MNRAS, 401, 1564-1569 \\ 
Sguera, V., Barlow, E.J., Bird, A.J., et al. 2005, A\&A, 444, 221-231 \\
Sguera, V., Bazzano, A., Bird, A. J., et al. 2006, ApJ, 646, 452-463 \\
Sguera, V., Ducci, L., Sidoli, L., Bazzano, A., Bassani, L., 2010, MNRAS, 402, L49-L53 \\
Sidoli, L., 2009,  AdSpR, 43, 1464-1470 \\
Sidoli, L., Paizis, A., \& Mereghetti, S., 2006, A\&A, 450, L9-L12 \\
Sidoli, L., Romano, P., Mereghetti, S., et al.,  2007, A\&A, 476, 1307-1315 \\
Sidoli, L., Romano, P., Mangano, V., et al., 2008, ApJ 687, 1230-1235 \\
Sidoli, L., Romano, P., Ducci, L., et al., 2009, MNRAS, 397, 1528-1538 \\
Sidoli, L., Esposito, P., Ducci, L., 2010, MNRAS, 409, 611-618  \\
Tagliaferri, G., Argan, A., Bellazzini, R., et al., 2010,  SPIE, 7732, 773217-773217-12 \\
Walter, R., Zurita Heras, J., Bassani, L., et al. 2006, A\&A, 453, 133-143 \\

%\begin{thebibliography}{}
%\end{thebibliography}

\end{document}